\pdfoutput=1

\documentclass[11pt,twoside,a4paper,cmspaper,final,collab]{cms-tdr}
\def\svnVersion{231468}\def\cmsCernNo{2013-179}\def\cmsCernDate{\today}\def\cmsMessage{Published in Physics Letters B as \href{http://dx.doi.org/10.1016/j.physletb.2014.02.033}{\doi{10.1016/j.physletb.2014.02.033.}}}

\begin{document}\cmsNoteHeader{B2G-12-023}

\hyphenation{had-ron-i-za-tion}
\hyphenation{cal-or-i-me-ter}
\hyphenation{de-vices}

\RCS$Revision: 225677 $
\RCS$HeadURL: svn+ssh://svn.cern.ch/reps/tdr2/papers/B2G-12-023/trunk/B2G-12-023.tex $
\RCS$Id: B2G-12-023.tex 225677 2014-02-03 15:15:28Z gbruno $
\newlength\cmsFigWidth
\ifthenelse{\boolean{cms@external}}{\setlength\cmsFigWidth{0.85\columnwidth}}{\setlength\cmsFigWidth{0.4\textwidth}}
\ifthenelse{\boolean{cms@external}}{\providecommand{\cmsLeft}{top}}{\providecommand{\cmsLeft}{left}}
\ifthenelse{\boolean{cms@external}}{\providecommand{\cmsRight}{bottom}}{\providecommand{\cmsRight}{right}}
\providecommand{\FEWZ}{\textsc{fewz}\xspace}
\cmsNoteHeader{B2G-12-023} 
\title{Search for baryon number violation in top-quark decays}

\date{\today}
\abstract{
A search for baryon number violation (BNV) in top-quark decays
is performed using pp collisions produced by the LHC at
$\sqrt{s}=8$\TeV.  The top-quark decay considered in this search
results in one light lepton (muon or electron), two jets, but no
neutrino in the final state. Data used for the analysis were collected by the
CMS detector and correspond to an integrated luminosity of 19.5\fbinv.
The event selection is optimized for top quarks
produced in pairs, with one undergoing the BNV decay
and the other the standard model hadronic decay to three jets. No
significant excess of events over the expected yield from standard model
processes is observed.
The upper limits at 95\% confidence level on the branching fraction of
the BNV top-quark decay are calculated
to be 0.0016 and 0.0017 for the muon and the electron channels,
respectively.
Assuming lepton universality, an upper limit of 0.0015 results from
the combination of the two channels.
These limits are the first that have been obtained on a BNV process
involving the top quark.}

\hypersetup{%
pdfauthor={CMS Collaboration},%
pdftitle={Search for baryon number violation in top quark decays},%
pdfsubject={CMS},%
pdfkeywords={LHC, CMS, top quark, baryon number}}

\maketitle 

\section{Introduction}
In the standard model (SM) of particle
physics~\cite{StandardModel67_1,StandardModel67_2,StandardModel67_3},
baryon number is a conserved quantity as a
consequence of an accidental symmetry of the Lagrangian. In fact, it has
been proven that extremely small violations can arise from
non-perturbative effects~\cite{PhysRevLett.37.8}.
Baryon number violation (BNV) is also predicted in several
scenarios of physics beyond the SM such as supersymmetry~\cite{Martin:1997ns,
Kane:2010zza}, grand unification~\cite{Georgi:1974sy}, and models with
black holes~\cite{Bekenstein:1971hc}.
Furthermore, BNV is a necessary condition for the observed asymmetry between
baryons and antibaryons in the Universe, assuming an evolution from a
symmetric initial state~\cite{Sakharov:1967dj}.

Despite these compelling reasons,
no direct evidence of BNV processes has been found to date.
Experiments have set stringent limits on BNV in
nucleon~\cite{PhysRevLett.102.141801}, $\tau$-lepton~\cite{Godang:1999ge,
Miyazaki:2005ng, tagkey201336}, c- and
b-hadrons~\cite{PhysRevD.79.097101, PhysRevD.83.091101, PhysRevD.84.072006},
and Z-boson~\cite{Abbiendi:1998nj} decays.
The possibility that BNV could occur in the decay of the top quark (t)
was first considered in Ref.~\cite{PhysRevD.72.095001}, in which a very
stringent bound of about $10^{-27}$  was derived on the branching
fraction of the decay $\cPqt \to \cPaqb
\cPaqc \ell^+$, where $\ell$ is either an electron or a muon, using
the experimental bound on the proton lifetime~\cite{PhysRevLett.102.141801}.
However, more recently it has been noted~\cite{Dong:2011rh} that cancellations
between different four-fermion interactions could allow much higher
rates of occurrence for the BNV decays
$\cPqt \to \cPaqb \cPaqc \Pgmp$
($\cPaqt \to \cPqb \cPqc \Pgmm$) and $\cPqt
\to \cPaqb \cPaqu \Pep$
($\cPaqt\to \cPqb \cPqu \Pem$).
Other BNV decays of the top quark, involving different flavors for the lepton
and quarks, are also discussed in Ref.~\cite{Dong:2011rh}, where in all
cases they are described as four-fermion effective interactions, in
which both baryon number and lepton number are violated.

In this Letter we search for evidence of such BNV top-quark decays
using $19.52 \pm 0.49\fbinv$ of pp collision data at $\sqrt{s} =
8$\TeV, collected in 2012 with the Compact Muon Solenoid (CMS)
detector~\cite{:2008zzk} at
the Large Hadron Collider (LHC). These decays are referred to as ``BNV
decays'' in the following, as opposed to the SM decay of the top quark
into a W boson and a down-type quark, the latter being a bottom quark
in about 99.8\% of the cases.
Assuming that the BNV decay branching fraction ($\mathcal{B}$) is ${\ll}1$,
the most suitable process for its observation
is expected to be pair production of top quarks
(\ttbar), where one top quark undergoes a SM decay into three
jets and the other the BNV decay. This process would have
the highest cross section among those involving at least one BNV decay
and could be effectively separated from background.
Two event selections, one for
the muon and one for the electron channel, are defined and optimized for such a
process. In both cases the final state consists of a lepton, five
quarks, and no neutrino.

\section{The CMS detector}
The central feature of the CMS apparatus is a $3.8\unit{T}$
superconducting solenoid of $6\unit{m}$ internal diameter.
Inside the coil are the silicon pixel and strip
tracker, the lead-tungstate crystal electromagnetic calorimeter (ECAL), and the
brass and scintillator hadron calorimeter. Muons are detected by four
layers of gas-ionization detectors embedded in the steel flux-return yoke. In
addition to the barrel and endcap detectors, CMS has extensive forward
calorimetry. A two-stage trigger system selects $\Pp\Pp$
collision events of interest for use in physics analyses.
CMS uses a right-handed coordinate system, with the origin at the
nominal interaction point, the $x$-axis pointing to the center of the
LHC ring, the $y$-axis pointing up (perpendicular to the LHC plane), and
the $z$-axis pointing along the counterclockwise-beam direction. The polar
angle $\theta$ is measured from the positive $z$-axis and the
azimuthal angle $\phi$ is measured in the $x$-$y$ plane.
Muons (electrons) are reconstructed and identified in the
pseudorapidity ($\eta= -\ln [\tan (\theta/2)]$) range  $\abs{\eta}<
2.4$ $(2.5)$.
The pixel (strip) tracker consists of three (ten) co-axial detection
layers in the central region and two (twelve) disk-shaped layers in
the forward region.
The inner tracker measures charged particle trajectories within the
pseudorapidity range $\abs{\eta} < 2.5$,
and provides an impact parameter resolution of $\sim$15\mum.
A detailed description of the CMS detector can be
found elsewhere~\cite{:2008zzk}.

\section{Trigger and datasets}

The data used for this analysis were collected using
isolated-lepton (muon or electron) plus multijet triggers.
In the muon (electron) trigger an isolated muon (electron) candidate
is required to have a transverse momentum \pt greater than 20
(25)\GeVc, $\abs{\eta} < 2.1\ (2.5)$, and be accompanied by at least
three jets in  $\abs{\eta} < 2.4$, with  $\pt >45, 35, 25\  (50, 40,
30)$\GeVc.
The trigger efficiency for signal
events passing the offline selection, described in
Section~\ref{sec:evselection}, is $83\pm 2 \%$ ($80\pm 2\%$) for the
muon (electron) analysis.

A number of simulated event samples with the most important
backgrounds are used to compare observations with SM expectations. The
$\mathrm{t\bar{t}}$+jets, W+jets, and Z+jets events are generated with
the \MADGRAPH v5.1.3.30 event generator~\cite{Alwall:2007st}. The
top-quark mass is set to 172.5\GeVcc and the branching fraction of
top-quark decays to a \PW\ boson and a b quark is assumed to be
one. \MADGRAPH is interfaced with \PYTHIA
v6.426~\cite{Sjostrand:2006za} to simulate parton fragmentation and
hadronization. For each production process,
data samples corresponding to 0, 1, 2, and 3 extra partons are merged
using the ``MLM'' matching prescription~\cite{Alwall:2008qv} in order to
yield a realistic spectrum of accompanying jets.
Diboson and QCD multijet events are generated with \PYTHIA, whereas
single-top-quark samples are generated with
\POWHEG~\cite{Nason:2004rx, Frixione:2007vw}.
Samples of {\ttbar}W and  {\ttbar}Z events with up to one extra
parton are generated  with \MADGRAPH.

A number of signal samples are generated with \MADGRAPH
v5.1.4.3~\cite{Alwall:2011uj} interfaced  with \PYTHIA
v8.165~\cite{Sjostrand2008852}.
Two of these samples correspond to
events with \ttbar-pair production in which one or both top quarks
have a BNV decay.
Three other simulated signal samples correspond to the tW, $t$-channel,
and $s$-channel processes giving rise to single top-quark production. In
each of these cases the top quark has a BNV decay, and samples
corresponding to zero and one extra parton are merged. All the
fermion-flavor-dependent parameters, which appear in the effective BNV
Lagrangian defined in~\cite{Dong:2011rh}, were set to unity. Different
choices for these values can in principle lead to variations in the
kinematical distributions of the top-quark decay products, but the
resulting impact on the final results of the search is negligible.

For all simulated samples, the hard interaction collision is
overlaid with a number of simulated minimum-bias collisions. The
resulting events are weighted to reproduce the distribution of the
number of inelastic collisions per bunch crossing (pileup) measured in
data.

A detailed simulation of particle propagation through the CMS
apparatus and detector response is performed with the \GEANTfour
v9.2~\cite{Agostinelli2003250, geant} toolkit.

\section{Event reconstruction and selection  \label{sec:evselection}}

The signal search is performed by counting events passing a ``tight''
selection. As explained in Section~\ref{sec:brmeasurement}, the
sensitivity of the search is substantially improved by also using a
``basic'' selection that includes the tight one.
These two event selections are described below.

\subsection{Basic selection}

Events are reconstructed using a particle-flow (PF)
algorithm~\cite{CMS-PAS-PFT-10-002}, which consists in reconstructing
and identifying each particle with an optimized combination of
all subdetector information.
Reconstructed particles are categorized into muons,
electrons, photons, charged hadrons and neutral hadrons.
At least one  primary vertex is required to be reconstructed in a
cylindrical region defined by the longitudinal distance $\abs{z} < 24$\unit{cm}
and radial distance $r < 2$\unit{cm} relative to the center of the CMS
detector.
The average number of reconstructed primary vertices per event is
approximately 15 for the 2012 data-taking period. The
reconstructed primary vertex with the largest $\sum \pt^{2}$ of all
associated tracks is assumed to be produced by the hard-scattering
process. All reconstructed muons, electrons and charged hadrons
used in this analysis are required to be associated with this primary
vertex.

Muons are identified by performing a combined  fit to position
measurements from both the inner tracker and the muon
detectors~\cite{Chatrchyan:2012xi}.
They are required to have $\pt > 25$\GeVc and $\abs{\eta} < 2.1$.
Their associated tracks are required to have measurements in at least
six of the inner tracker layers, including at least one pixel detector
layer, a combined fit $\chi^{2}$ per degree of freedom smaller
than 10,
and to be reconstructed using at least two muon detector layers.
In addition, the transverse (longitudinal) impact parameter of the
muon track relative to the reconstructed primary vertex is required
to be smaller than 0.2\unit{cm} (0.5\unit{cm}).

Electrons are identified~\cite{Chatrchyan:2013dga}
as tracks reconstructed in the inner tracker with measured momenta
compatible with their associated energy depositions in the ECAL\@.
Electrons  are required to have $\pt > 30$\GeVc and $\abs{\eta} < 2.5$, with
the exclusion of the transition region between barrel and endcaps
defined by $1.444<\abs{\eta} < 1.566$.
The transverse (longitudinal) impact parameter of the
electron track relative to the reconstructed primary vertex is required
to be smaller than 0.02\unit{cm} (0.1\unit{cm}).  These requirements are tighter
than in the case of muons in order to reject electrons originating from photon
conversions, and misidentified hadrons coming from pileup collisions.
Additional photon conversion rejection
requirements~\cite{Chatrchyan:2013dga} are also applied.

Muon and electron candidates are required to be isolated. Isolation is
defined via the variable
\begin{equation}
I_{\text{rel}}^{\ell} = \frac{E_{\text{ch}} + E_{\text{nh}} + E_{\gamma}}{\pt^{\ell}c},
\label{eq:iso}
\end{equation}
where $\pt^{\ell}$ is the lepton transverse momentum,
$E_{\text{ch}}$ is the transverse energy deposited by charged
hadrons in a cone with aperture $\Delta R = 0.4$ (0.3) in $(\eta,\phi)$ around
the muon (electron) track, and $E_{\text{nh}}$ ($E_{\gamma}$)
is the transverse energy of neutral hadrons (photons) within this
cone. The transverse energies in Eq.~(\ref{eq:iso}) are defined as the
scalar sum of the transverse momenta of all contributing
particles. Muon (electron) candidates are required to have
$I_{\text{rel}}^{\ell} < 0.12$ (0.10).

Events with exactly one lepton candidate satisfying the
above criteria are selected for further consideration. Events with
one or more additional reconstructed muons (electrons)
with \pt $>$ 10 (15)\GeVc, $\abs{\eta}<2.5$, $I_{\text{rel}}^{\ell} <
0.2$ are rejected.

All the particles identified by the PF algorithm that are
associated with the primary vertex are clustered into jets using the
anti-\kt algorithm~\cite{antikt} with a distance parameter of 0.5.
Corrections to the jet energy scale are applied to account for the
dependence of the detector response to jets as a function of $\eta$
and \pt and the effects of pileup~\cite{Chatrchyan:2011ds}.
The energy of reconstructed jets in simulated events is also smeared to
account for the 5--10\% discrepancy in energy resolution that is
observed between data and simulation~\cite{Chatrchyan:2011ds}.
At least five jets are
required with  $\pt> 70,$ 55, 40, 30,
30\GeVc and $\abs{\eta}<2.4$. The offline jet \pt thresholds are chosen
such that all selected jets are in the trigger efficiency plateau.
In addition, at least one of these jets must be identified as
originating from a b quark (``b tagging'')
by the ``combined secondary-vertex'' (CSV)
algorithm tuned for high efficiency~\cite{Chatrchyan:2012jua}. This
algorithm combines information about the impact parameter of tracks
and reconstructed secondary vertices within the jet. Its
typical efficiency for tagging b-quark jets is about 80\%,
whereas the mistagging efficiency  is about 10\% for jets produced by
the hadronization of light quarks (u, d, s) or gluons, and about 35\% for
jets from c quarks~\cite{Chatrchyan:2012jua}.

\subsection{Tight selection}

\begin{figure}[htb]
\begin{center}
\includegraphics[width=0.48\textwidth]{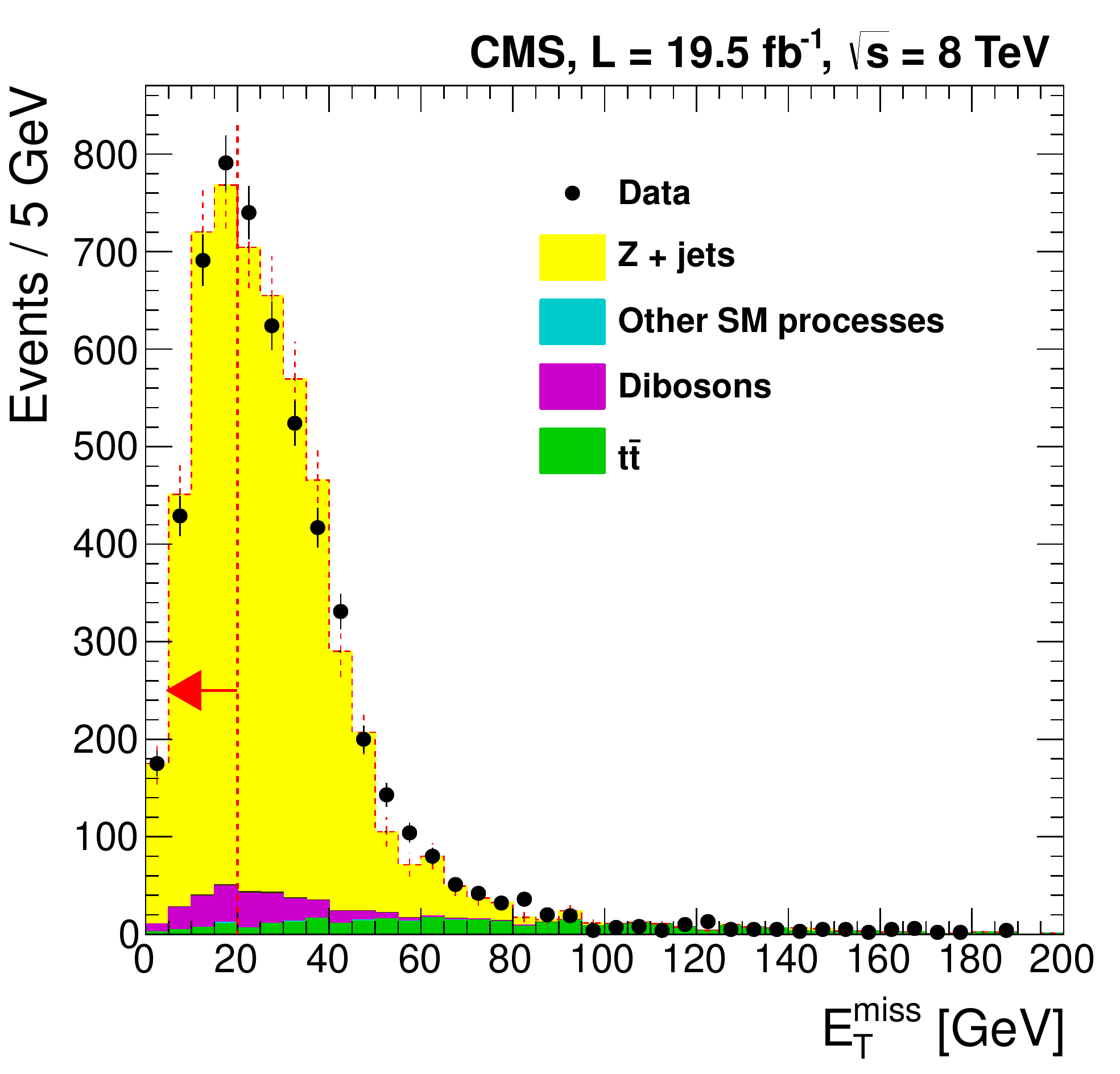}
\caption{Distributions of \MET in both data and simulation for a sample
of events enriched in $\cPZ+4\,\text{jets} \to\Pgmp \Pgmm+4\,\text{jets}$
events. The simulated distribution is normalized to that observed in
data. For both data and simulation the error bars indicate the
statistical uncertainty. The vertical dashed line and the arrow
indicate the \MET signal region used in this analysis. The
total contribution from other SM processes, such as single-top-quark,
W+jets, {\ttbar}W, and {\ttbar}Z productions is very small and
is hardly visible in the plot.}
\label{fig:metZ}
\end{center}
\end{figure}

Two additional requirements define the tight
event selection used for the signal search.
The first is the presence of small missing transverse energy (\MET).
In PF reconstruction, \MET is defined as the
modulus
of the vector sum of the transverse momenta of all
reconstructed particles (charged and neutral) in the
event.
The jet energy scale corrections are also used to correct the \MET
value~\cite{Chatrchyan:2011tn}.
In order to pass the tight selection, events are required to have $\MET
<20\GeV$.
The validity of the simulation for low-\MET events
is verified using a data sample enriched in $\cPZ+$4 jets
$\to$\Pgmp \Pgmm$+$4 jets events.
Events in this sample are required to have at least four jets with
$\pt>30$\GeVc and $\abs{\eta}<2.4$ in addition to two muons with $\pt>20$\GeVc,
$\abs{\eta}<2.1$, $I_{\text{rel}}^{\ell} < 0.10$, and invariant mass in
the range 76--104\GeVcc.
The \MET distributions obtained in data
and simulation for this event sample are shown in
Fig.~\ref{fig:metZ}. The simulated distribution is
normalized to that observed in data and the two agree within
statistical uncertainties. The disagreement in overall yield before
normalization is at the level of 7\%, which is covered by statistical and
theoretical~\cite{PhysRevD.85.031501} uncertainties. Very similar
results are obtained with a di-electron data sample or by requiring five
additional jets, instead of four, in either the muon or the electron
data samples. With five additional jets, however, the statistical uncertainties
are significantly larger.

The second requirement for an event to pass the tight selection is the
compatibility of its kinematic properties with the final state
produced by \ttbar events where one top quark
has a fully hadronic SM decay and the other the BNV decay. This
compatibility is tested using the following variable:
\begin{equation}
\chi^2= \sum_i \cfrac{(x_i -\bar{x_i})^2}{\sigma_i^2},
\label{eq:chi2}
\end{equation}
where the $x_i$ are the reconstructed invariant mass of
the W boson from the hadronically decaying top quark, the reconstructed
invariant mass of the hadronically decaying top quark, and  the reconstructed
invariant mass of the top quark with the BNV decay. The values of
$\bar{x_i}$ and $\sigma_i$ are the expectation values and standard
deviations of Gaussian fits to the $x_i$ distributions obtained from
simulated \ttbar events with the two top quarks undergoing the BNV and
the fully hadronic SM decay, respectively.
Simulation information is used to obtain the correct jet-to-parton
association. The number of jet combinations is
reduced by not associating jets tagged by the CSV algorithm
with the W decay. All
other combinations in the event are considered and the one with the
smallest $\chi^2$ is retained. For signal events,
the correct jet combination is expected to be chosen in about 60\% of
the cases.
In the tight selection, the smallest $\chi^2$ is required to be
less than 20.

The values of the thresholds on the lepton \pt,
\MET, $\chi^2$, as well as the configuration of the b-tagging
algorithm were chosen by minimizing the expected upper limit on $\mathcal{B}$ at 95\% confidence level (CL).
This procedure also retains high sensitivity for an observation of a BNV decay.

\section{Signal search strategy \label{sec:brmeasurement}}

The search proceeds in the following way: for each assumed value of
$\mathcal{B}$,
the expected contributions from \ttbar
($N^{\mathrm{B}}_{\ttbar}$) and tW
($N^{\mathrm{B}}_{\cPqt\PW}$) production
to the yield in the basic selection are scaled such that the total
expected yield is normalized to the observed number of events
($N^\mathrm{B}_{\text{obs}}$). The sum of the \ttbar
and tW yields in the tight selection ($N^\mathrm{T}_{\text{top}}$)
is then extracted using
the efficiencies,
$\epsilon^{(\mathrm{T|B})}_{\ttbar}$ and
$\epsilon^{(\mathrm{T|B})}_{\cPqt\PW}$, to pass the tight selection for
\ttbar and tW events that satisfy the basic selection
criteria. Finally, the comparison between the total expected and
observed numbers of events in the tight selection, which is
significantly more efficient for the signal than for the background,
is used to infer the presence of a signal or set an upper limit on
$\mathcal{B}$.
The impact of a number of systematic uncertainties is significantly
reduced as a result of the normalization of simulation to data in the
basic selection.
Indeed,  using this approach, the expected upper limit at 95\% CL on
$\mathcal{B}$ is found to improve by a factor of 2.5, while the expected
significance of a signal-like deviation from SM expectations increases
from about 1.2 standard deviations to 3.6 for a signal with
$\mathcal{B}=0.005$.
In this procedure we neglect the contributions to a possible BNV
signal from events with single-top-quark production via $s$- and
$t$-channels, {\ttbar}W, and {\ttbar}Z, which are treated as
non-top background. These contributions are expected to be negligible,
as can be inferred from the yields of these processes given in
Tables~\ref{tab::basictightnoBR_Wttmeasured}
and~\ref{tab::basictightnoBR_WttmeasuredEle}, as estimated from simulation.

Following the approach outlined above, the expression for the expected
total yield in the tight selection ($N^\mathrm{T}_{\text{exp}}$) is:
\ifthenelse{\boolean{cms@external}}{
\begin{multline}
N^\mathrm{T}_{\text{exp}}= N^\mathrm{T}_{\text{top}} + N^\mathrm{T}_{\text{bck}} = \left ( N^\mathrm{B}_{\text{obs}} - N^\mathrm{B}_{\text{bck}} \right ) \Big[
\frac{N^{\mathrm{B}}_{\ttbar}}{N^{\mathrm{B}}_{\ttbar}+N^{\mathrm{B}}_{\cPqt\PW}}\cdot
\epsilon^{(\mathrm{T|B})}_{\ttbar} +\\
\frac{N^{\mathrm{B}}_{\cPqt\PW}}{N^{\mathrm{B}}_{\ttbar}+N^{\mathrm{B}}_{\cPqt\PW}}\cdot
\epsilon^{(\mathrm{T|B})}_{\cPqt\PW}
\Big] + N^\mathrm{T}_{\text{bck}},
\label{eq:BR0}
\end{multline}
}{
\begin{equation}
N^\mathrm{T}_{\text{exp}}= N^\mathrm{T}_{\text{top}} + N^\mathrm{T}_{\text{bck}} = \left ( N^\mathrm{B}_{\text{obs}} - N^\mathrm{B}_{\text{bck}} \right ) \left [
\frac{N^{\mathrm{B}}_{\ttbar}}{N^{\mathrm{B}}_{\ttbar}+N^{\mathrm{B}}_{\cPqt\PW}}\cdot
\epsilon^{(\mathrm{T|B})}_{\ttbar} +
\frac{N^{\mathrm{B}}_{\cPqt\PW}}{N^{\mathrm{B}}_{\ttbar}+N^{\mathrm{B}}_{\cPqt\PW}}\cdot
\epsilon^{(\mathrm{T|B})}_{\cPqt\PW}
\right ] + N^\mathrm{T}_{\text{bck}},
\label{eq:BR0}
\end{equation}
}
where
$N^\mathrm{B}_{\text{bck}}$ ($N^\mathrm{T}_{\text{bck}}$) is the yield of
the non-top background (including $s$- and $t$-channel single-top-quark,
{\ttbar}W, and {\ttbar}Z production events, as discussed
above) in the basic (tight) selection.

All the quantities in the square brackets of
Eq.~(\ref{eq:BR0}) are functions of $\mathcal{B}$ and can be expressed in
terms of the \ttbar and tW cross sections ($\sigma_{\ttbar}$ and
$\sigma_{\cPqt\PW}$, respectively), integrated luminosity, and
efficiencies for passing the basic and tight selections.
In terms of these variables, Eq.~(\ref{eq:BR0}) becomes:
\ifthenelse{\boolean{cms@external}}{
\begin{multline}
N^\mathrm{T}_{\text{exp}}
= \left( N^\mathrm{B}_{\text{obs}} - N^\mathrm{B}_{\text{bck}} \right ) \Big [
\frac{1}{1+\frac{\sigma_{\cPqt\PW}\epsilon^{\mathrm{B}}_{\cPqt\PW}(\mathcal{B})}{\sigma_{\ttbar}\epsilon^{\mathrm{B}}_{\ttbar}(\mathcal{B})}}\cdot
\frac{\epsilon^{\mathrm{T}}_{\ttbar}(\mathcal{B})}{\epsilon^{\mathrm{B}}_{\ttbar}(\mathcal{B})} +\\
\frac{1}{1+\frac{\sigma_{\ttbar}\epsilon^{\mathrm{B}}_{\ttbar}(\mathcal{B})}{\sigma_{\cPqt\PW}\epsilon^{\mathrm{B}}_{\cPqt\PW}(\mathcal{B})}}\cdot
\frac{\epsilon^{\mathrm{T}}_{\cPqt\PW}(\mathcal{B})}{\epsilon^{\mathrm{B}}_{\cPqt\PW}(\mathcal{B})}
\Big ] + N^\mathrm{T}_{\text{bck}}.
\label{eq:BR}
\end{multline}
}{
\begin{equation}
N^\mathrm{T}_{\text{exp}}
= \left ( N^\mathrm{B}_{\text{obs}} - N^\mathrm{B}_{\text{bck}} \right ) \left [
\frac{1}{1+\frac{\sigma_{\cPqt\PW}\epsilon^{\mathrm{B}}_{\cPqt\PW}(\mathcal{B})}{\sigma_{\ttbar}\epsilon^{\mathrm{B}}_{\ttbar}(\mathcal{B})}}\cdot
\frac{\epsilon^{\mathrm{T}}_{\ttbar}(\mathcal{B})}{\epsilon^{\mathrm{B}}_{\ttbar}(\mathcal{B})} +
\frac{1}{1+\frac{\sigma_{\ttbar}\epsilon^{\mathrm{B}}_{\ttbar}(\mathcal{B})}{\sigma_{\cPqt\PW}\epsilon^{\mathrm{B}}_{\cPqt\PW}(\mathcal{B})}}\cdot
\frac{\epsilon^{\mathrm{T}}_{\cPqt\PW}(\mathcal{B})}{\epsilon^{\mathrm{B}}_{\cPqt\PW}(\mathcal{B})}
\right ] + N^\mathrm{T}_{\text{bck}}.
\label{eq:BR}
\end{equation}
}
In Eq.~(\ref{eq:BR}), $\epsilon^\mathrm{B}_{\ttbar}$
($\epsilon^\mathrm{T}_{\ttbar}$)
indicates the efficiency of the basic (tight) selection for \ttbar
events. Similarly, $\epsilon^\mathrm{B}_{\cPqt\PW}$
($\epsilon^\mathrm{T}_{\cPqt\PW}$) indicates the
efficiency of the basic (tight) selection for tW events. Each of these
four efficiency values is a function of $\mathcal{B}$ and of three (for \ttbar
events) or two (for tW events) efficiency values,
which correspond to the different decay modes:
\begin{equation}
\epsilon^\mathrm{X}_{\ttbar}(\mathcal{B}) = 2\mathcal{B}(1-\mathcal{B})\epsilon^\mathrm{X}_{\mathrm{BNV,\, SM}} + (1-\mathcal{B})^2\epsilon^\mathrm{X}_{\mathrm{SM,\, SM}} + \mathcal{B}^2\epsilon^\mathrm{X}_{\mathrm{BNV,\, BNV}},
\label{eq:effBR}
\end{equation}
\begin{equation}
\epsilon^\mathrm{X}_{\cPqt\PW}(\mathcal{B}) = (1-\mathcal{B})\epsilon^\mathrm{X}_{\mathrm{SM}} + \mathcal{B}\epsilon^\mathrm{X}_{\mathrm{BNV}},
\label{eq:effBR2}
\end{equation}
where X = B, T with B and T denoting the basic and tight selection,
respectively, and SM (BNV) indicating the SM (BNV) decay mode of
the top quark.
With the adopted approach,
the search is mostly sensitive to uncertainties in the ratio of
$\epsilon^\mathrm{T}_{\mathrm{SM,\,SM}}$ to
$\epsilon^\mathrm{B}_{\mathrm{SM,\,SM}}$, $N^\mathrm{B}_{\text{bck}}$
and $N^\mathrm{T}_{\text{bck}}$.

\section{Background evaluation \label{sec:bgestimate}}

In order to evaluate the expected yield in the tight selection
(Eq.~(\ref{eq:BR})), a number of backgrounds need to be
estimated.

\subsection{Top-quark and electroweak backgrounds \label{sec:bgestimate1} }

The main background in this analysis is from \ttbar events where one of
the two top quarks decays to a lepton, a neutrino, and a b quark while the
other one decays to three quarks. As described in
Section~\ref{sec:brmeasurement}, the estimates of the \ttbar and tW
yields require knowledge of the efficiencies for \ttbar and tW events, which
satisfy the basic selection criteria, to also pass the tight selection. These
efficiencies are obtained from simulation.
The required \ttbar cross section is the prediction at
next-to-next-to-leading order (NNLO) that includes soft-gluon resummation at
next-to-next-to-leading log (NNLL)~\cite{PhysRevLett.110.252004}. 
The tW cross section is the approximate NNLO prediction from
soft-gluon resummation at NNLL~\cite{xsec_ttbar}.

The second-largest background  is
represented by W and Z production in association with jets.
The theoretical predictions for the
$\PW+\text{jets} \to \ell\nu +\text{jets}$ and
$\cPZ/\gamma^*+ \text{jets}
\to \ell\ell + \text{jets}$ processes, where $\ell$ indicates a lepton,
are computed by {\FEWZ}~\cite{PhysRevLett.96.231803, Gavin:2010az} at
next-to-next-to-leading order.
The efficiencies for events produced by these processes
to pass the basic and tight selections are evaluated from simulation and, using
the measured value of the integrated
luminosity~\cite{CMS-PAS-LUM-13-001}, the yields in the basic and
tight selections are obtained.

The contributions to the yield in the basic and tight selections from
single-top-quark production via $s$- and $t$-channel processes, and from
WW, WZ, ZZ, {\ttbar}W, and {\ttbar}Z production, are also evaluated
from simulation. The cross section value for single-top-quark production via
$s$-channel is computed using next-to-next-to-leading-logarithm
resummation of soft and collinear
gluon corrections~\cite{Kidonakis:2010tc}. The cross section
values for {\ttbar}W and {\ttbar}Z are computed
at leading-order (LO) as provided by \MADGRAPH, including the
contributions at LO from processes yielding one
extra jet. In all other cases the next-to-leading order (NLO)
theoretical predictions, as obtained from
{\MCFM}~\cite{Campbell201010}, are used. The sum of yields
predicted by the simulation for all these processes is less than 1\%
of the total expected yield in both the basic and tight
selections.

All cross section values used in the analysis are listed in
the second column of Tables~\ref{tab::basictightnoBR_Wttmeasured}
and~\ref{tab::basictightnoBR_WttmeasuredEle}. The yields reported in
these tables are discussed in Section~\ref{sec:results}.

\begin{table*}[htbp]
\begin{center}
\topcaption{\label{tab::basictightnoBR_Wttmeasured} Muon channel:
assumed cross section values, expected (as discussed in
Section~\ref{sec:bgestimate})  and observed
yields in the basic and tight selections for an
assumed $\mathcal{B}$ value of zero.
The ``Basic'' and ``Corrected basic'' columns
report the yields in the basic selection before and after
the normalization procedure described in
Section~\ref{sec:brmeasurement}.
The uncertainties include both statistical and systematic
contributions. Many of the reported uncertainties are either
correlated or anticorrelated, which explains why the uncertainties in
the total expected yields are smaller than those in some of their
components. By construction the total expected yield after
the normalization procedure is equal to the observed yield.}
\begin{tabular}{ l | c  c  c  c }

Process & Cross section (pb) & Basic & Corrected basic & Tight \\
\hline
\ttbar  &  246  & 38800 $\pm$ 7800 & 38700 $\pm$ 3600 & 2210 $\pm$ 220 \\
W+jets  & 37500  &  \multicolumn{2}{c}{6300 $\pm$ 3200} & 230 $\pm$ 120 \\
Z+jets  &  3500  & \multicolumn{2}{c}{380 $\pm$ 190} & 32 $\pm$ 18 \\
tW  &  22.2  & 1160 $\pm$ 180 & 1160 $\pm$ 220 & 49 $\pm$ 9 \\
$t$-channel  &  87.1  & \multicolumn{2}{c}{250 $\pm$ 130} & 5.7 $\pm$ 3.0 \\
$s$-channel  &  5.55  & \multicolumn{2}{c}{31 $\pm$ 16} & 0.84 $\pm$ 0.52 \\
WW  &  54.8  & \multicolumn{2}{c}{86 $\pm$ 43} & 3.1 $\pm$ 1.7 \\
WZ  &  33.2  & \multicolumn{2}{c}{41 $\pm$ 21} & 1.43 $\pm$ 0.78 \\
ZZ  &  17.7  & \multicolumn{2}{c}{5.5 $\pm$ 2.8} & 0.49 $\pm$ 0.28 \\
\ttbar\PW &  0.23  & \multicolumn{2}{c}{128 $\pm$ 64} & 5.9 $\pm$
3.0 \\
{\ttbar}Z  &  0.17  & \multicolumn{2}{c}{79 $\pm$ 40} & 4.1 $\pm$
2.1 \\
QCD  &  ---  & \multicolumn{2}{c}{760 $\pm$ 530} & 112 $\pm$ 56 \\\hline
Total exp. & --- & 48000 $\pm$ 8600 & 47950 $\pm$  220 & 2650 $\pm$  130 \\
Data & --- & \multicolumn{2}{c}{47951} & 2614 \\
\end{tabular}
\end{center}
\end{table*}

\begin{table*}[htbp]
\begin{center}
\topcaption{\label{tab::basictightnoBR_WttmeasuredEle} Electron channel:
assumed cross section values, expected (as discussed in
Section~\ref{sec:bgestimate})  and observed
yields in the basic and tight selections for an
assumed $\mathcal{B}$ value of zero.
The ``Basic'' and ``Corrected basic'' columns
report the yields in the basic selection before and after
the normalization procedure described in
Section~\ref{sec:brmeasurement}.
The uncertainties include both statistical and systematic
contributions. Many of the reported uncertainties are either
correlated or anticorrelated, which explains why the uncertainties in
the total expected yields are smaller than those in some of their
components. By construction the total expected yield after
the normalization procedure is equal to the observed yield.
}
\begin{tabular}{l|cccc}

Process & Cross section (pb) & Basic & Corrected basic & Tight  \\
\hline
\ttbar &  246  & 38200 $\pm$ 7700 & 38400 $\pm$ 3700 & 2040 $\pm$ 220
\\
W+jets  &  37500  & \multicolumn{2}{c}{6500 $\pm$ 3300} & 240 $\pm$ 120 \\
Z+jets   &  3500  & \multicolumn{2}{c}{760 $\pm$ 380} & 85 $\pm$ 45 \\
tW  &  22.2  & 1110 $\pm$ 170 & 1120 $\pm$ 210 & 35.6 $\pm$ 6.3 \\
$t$-channel  &  87.1  & \multicolumn{2}{c}{230 $\pm$ 120} & 6.6 $\pm$ 3.6 \\
$s$-channel  &  5.55  & \multicolumn{2}{c}{27 $\pm$ 14} & 0.70 $\pm$ 0.50 \\
WW  &  54.8  & \multicolumn{2}{c}{78 $\pm$ 39} & 3.7 $\pm$ 2.0 \\
WZ  &  33.2  & \multicolumn{2}{c}{45 $\pm$ 23} & 2.1 $\pm$ 1.1 \\
ZZ  &  17.7  & \multicolumn{2}{c}{11.0 $\pm$ 5.6} & 1.40 $\pm$ 0.70 \\
{\ttbar}W  &  0.23  & \multicolumn{2}{c}{132 $\pm$ 66} & 6.2 $\pm$ 3.1 \\
{\ttbar}Z  &  0.17  & \multicolumn{2}{c}{86 $\pm$ 43} & 4.4 $\pm$ 2.2 \\
QCD  &  ---  & \multicolumn{2}{c}{2800 $\pm$ 1400} & 330 $\pm$ 170 \\
\hline
Total exp.  &  ---  & 50000 $\pm$ 9300 & 50110 $\pm$ 220 & 2750 $\pm$ 160  \\
Data  & --- & \multicolumn{2}{c}{50108}  & 2703 \\
\end{tabular}
\end{center}
\end{table*}

\subsection{QCD multijet background \label{sec:bgestimate2}}

The QCD multijet background yields are evaluated with two methods, depending
on the event selection.

In the first method  the isolation requirements for the
leptons are inverted (becoming $0.12 < I_{\text{rel}}^{\ell} < 0.2$
for muons and $0.10 <I_{\text{rel}}^{\ell} < 0.2$ for electrons) in
order to enhance the presence of QCD multijet events. These selections
are denoted as anti-isolated basic and anti-isolated tight in the
following, as opposed to the (isolated) basic and tight selections
used in the analysis.
The yield of the QCD multijet background in either the tight or basic
selection ($N_{\mathrm{QCD}}$) can thus be inferred using the
following equation:
\begin{equation}
N_{\mathrm{QCD}} = R \cdot (N^{\text{antiiso}}_{\text{data}} - N^{\text{antiiso}}_{\text{nonQCD}}),
\label{QCDestimate1}
\end{equation}
where $R$ is the ratio of the numbers of
QCD multijet events in the isolated and anti-isolated selections,
$N^{\text{antiiso}}_{\text{data}}$ is the yield observed in data
in the anti-isolated selection, and
$N^{\text{antiiso}}_{\text{nonQCD}}$ is the contribution in the
anti-isolated selection from other SM processes.
The value of $N^{\text{antiiso}}_{\text{nonQCD}}$ is estimated
from simulation using the cross section values discussed in
Section~\ref{sec:bgestimate1}. The value of $R$
is estimated
from data using the approximation $R=f/(1-f)$, where $f$ is the so-called
``misidentification rate''. The misidentification rate is defined as
the probability that a genuine jet that has passed all
lepton-identification criteria and a looser isolation threshold
($I_{\text{rel}}^{\ell} < 0.2$)
also passes the final analysis isolation threshold. The value of $f$
is obtained from data in five lepton \pt bins using a sample of events
enriched in $\cPZ+\text{jets}\to\Pgmp \Pgmm+\text{jets}$ events, where
a third, loosely isolated lepton, a muon or an electron, is also found.
The estimate of the QCD multijet yield is then determined using
Eq.~(\ref{QCDestimate1}) in each lepton \pt bin.
The misidentification rate is measured with events whose topology is
different from that of events in the final selection.  This difference
gives rise to the dominant uncertainty in the estimation of the
misidentification rate, which is assessed to be 20\% from the
difference observed in simulation between the true and predicted yields.
The systematic uncertainty
in  $N^{\text{antiiso}}_{\text{nonQCD}}$ is in the range 20--25\%
depending on the selection. After also taking into account the
statistical uncertainties,
this results in a 50
In the case of the muon basic selection the systematic uncertainty in
$N^{\text{antiiso}}_{\text{nonQCD}}$ is larger than the
difference $N^{\text{antiiso}}_{\text{data}} -
N^{\text{antiiso}}_{\text{nonQCD}}$ and therefore prevents a sufficiently
accurate estimate of the QCD multijet yield. For this reason, a
second method, which is described below, is used for this
specific selection.
In the electron analysis the uncertainties in the QCD multijet
background estimates in the basic and tight selections are treated as
fully correlated.
The numbers relevant for the QCD multijet yield
estimation with this first method are given in Table~\ref{QCDestimate}.
\begin{table*}
\begin{center}
\topcaption{\label{QCDestimate} Muon and electron channels: numbers
relevant for the estimate of the QCD multijet yield based on the
misidentification rate measurement (Eq.~(\ref{QCDestimate1})).
Only the average value of $f$ is reported, while values computed in
bins of \pt are used in the analysis.}
\begin{tabular}{ l | c  c  c  c  }
\multicolumn{5}{c}{Muon channel} \\ \hline
Selection &  $N^{\text{antiiso}}_{\text{data}}$  &
$N^{\text{antiiso}}_{\text{nonQCD}}$  & $f$ & $N_{\mathrm{QCD}}$ \\
\hline
Tight & 412 & 268 $\pm$ 55 & 0.44 $\pm$ 0.09 & 112 $\pm$ 56 \\
\hline
\noalign{\vskip 2ex}
\multicolumn{5}{c}{Electron channel} \\
\hline
Selection &  $N^{\text{antiiso}}_{\text{data}}$  &
$N^{\text{antiiso}}_{\text{nonQCD}}$  & $f$ & $N_{\mathrm{QCD}}$ \\
\hline
Basic & 7162  & 4600 $\pm$ 900 & $0.51 \pm 0.10$  & 2800 $\pm$ 1400 \\
Tight & 542 & 230 $\pm$ 48 & $0.51 \pm 0.10$   & 330 $\pm$ 170 \\
\end{tabular}
\end{center}
\end{table*}

In the second method the assumption is made that \MET and
$\chi^2$ are not correlated for QCD multijet events, and the estimated
QCD multijet yield in the basic selection ($N^{\mathrm{B}}_{\mathrm{QCD}}$) is
obtained from the following equation:
\begin{equation}
N^{\mathrm{B}}_{\mathrm{QCD}} =\frac{N^{\mathrm{T}}_{\mathrm{QCD}}}{\epsilon_{\MET}\epsilon_{\chi^2}},
\label{QCDestimate2}
\end{equation}
where $N^{\mathrm{T}}_{\mathrm{QCD}}$ is the QCD multijet yield in the
tight selection, as obtained  with the first method, and
$\epsilon_{\MET}$ ($\epsilon_{\chi^2}$) is the probability that
a QCD multijet event that passes the basic selection has $\MET<20\GeV$
($\chi^2<20$).
The values of $\epsilon_{\MET}$ and $\epsilon_{\chi^2}$ are taken from
simulation. A total uncertainty of 50\% in the product of
$\epsilon_{\MET}$ and $\epsilon_{\chi^2}$ is assumed, which yields, together
with the 50\% uncertainty in $N^{\mathrm{T}}_{\mathrm{QCD}}$, an
overall 70\%  uncertainty in the estimate of the QCD multijet
background in the muon basic selection.
The partial correlation with the uncertainty in the
QCD multijet background estimate in the muon tight selection is taken
into account when determining the final results of the analysis.
The numbers relevant for the QCD multijet yield estimation
in the muon basic selection with this second method are given in
Table~\ref{QCDestimateBASICMU}.
\begin{table}
\begin{center}
\topcaption{\label{QCDestimateBASICMU} Muon channel:  numbers relevant
for the estimate of the QCD multijet yield based on
Eq.~(\ref{QCDestimate2}). As stated in the text, this method is only
used for the muon basic selection.
}
\begin{tabular}{ l | c  c  c  }
\multicolumn{4}{c}{Muon channel} \\ \hline
Selection & $\epsilon_{\MET}$  & $\epsilon_{\chi^2}$  &  $N^{\mathrm{B}}_{\mathrm{QCD}}$ \\
\hline
Basic & $ 0.33 \pm 0.12 $  & $ 0.45 \pm 0.16$  & $ 760 \pm 530  $
\end{tabular}
\end{center}
\end{table}

In the electron analysis the contribution of $\gamma$+jets
processes can potentially give rise to events that pass the basic and
the tight selections. The isolated photon in the event can convert before
reaching the calorimeters and be identified as a single isolated
electron. From simulation studies it turns out that the
central values of the QCD multijet yields estimated with the first method in
both the basic and tight selections need to be increased by $2\%$ to
account for this contribution.

\section{Systematic uncertainties \label{sec:systematic}}

The observable of the likelihood function used in the analysis is the
yield in data for the tight selection ($N^\mathrm{T}_\text{obs}$), while
the parameter of interest is $\mathcal{B}$. A number of other quantities
appear in the likelihood function and
affect the estimate of $N^\mathrm{T}_{\text{exp}}$.
They are $N^\mathrm{B}_{\text{bck}}$,
$N^\mathrm{T}_{\text{bck}}$, the ratio
$\sigma_{\cPqt\PW}/\sigma_{\ttbar}$, and the ten efficiencies in
Eqs.~(\ref{eq:BR}-\ref{eq:effBR2}).
They are estimated as
described in Section~\ref{sec:bgestimate}.
Many of these quantities are correlated because of
common sources of systematic uncertainties. These correlations are handled
using the method presented in Ref.~\cite{morphing}, where the $j$-th
source of systematic uncertainty is associated with a nuisance
parameter of true value $u_j$  constrained by a normal
probability density function (PDF) $\mathcal{G}(u_j)$. The method
results in the parameterization, $\theta_i({u_j})$, of the $i$-th
likelihood quantity $\theta_i$  in terms of all the $u_j$ nuisance parameters.
Other quantities in the likelihood function are instead
either assumed to be independent of any other ($\sigma_{\cPqt\PW}$,
$\sigma_{\ttbar}$, and $N^\mathrm{B}_{\text{obs}}$) or
correlated with a single other quantity (the QCD multijet contributions
to $N^\mathrm{B}_{\text{bck}}$ and $N^\mathrm{T}_{\text{bck}}$).
In these cases the quantities are simply
constrained in the likelihood function by a
lognormal PDF $\rho_k(\tilde \theta_k | \theta_k )$,
which describes the probability of measuring a value $\tilde
\theta_k$ for the $k$-th likelihood quantity should its true
value be $\theta_k$, and which takes into account possible correlations.
With this approach, the likelihood function $\mathcal{L}$
reads:
\ifthenelse{\boolean{cms@external}}{
\begin{multline}
\label{eq:Likelihood}
\mathcal{L}(  N^\mathrm{T}_\text{obs} \, | \,  {\mathcal B} , {\theta_i({u_j})}, \theta_k ) =
\mathcal{P}\!\left(  N^\mathrm{T}_\text{obs} \, | \,
N^\mathrm{T}_{\text{exp}} ({\mathcal B}, {\theta_i({u_j})} ,
\theta_k ) \right)  \cdot\\
\prod_j \mathcal{G}(u_j) \cdot \prod_k \rho(\tilde{\theta}_k  \, |  \, \theta_k ),
\end{multline}
}{
\begin{equation}
\label{eq:Likelihood}
\mathcal{L}(  N^\mathrm{T}_\text{obs} \, | \,  {\mathcal B} , {\theta_i({u_j})}, \theta_k ) =
\mathcal{P}\!\left(  N^\mathrm{T}_\text{obs} \, | \,
N^\mathrm{T}_{\text{exp}} ({\mathcal B}, {\theta_i({u_j})} ,
\theta_k ) \right)  \cdot \prod_j \mathcal{G}(u_j) \cdot \prod_k \rho(\tilde{\theta}_k  \, |  \, \theta_k ),
\end{equation}
}
where $\mathcal{P}\!\left(  N^\mathrm{T}_{obs} \, | \, N^\mathrm{T}_{\text{exp}} ({\mathcal B},
{\theta_i({u_j})},  \theta_k) \right)$  indicates the Poisson PDF evaluated at
$N^\mathrm{T}_\text{obs}$ and with
expectation value $N^\mathrm{T}_{\text{exp}} ({\mathcal B},
{\theta_i({u_j})},  \theta_k)$ given by Eq.~(\ref{eq:BR}).

The sources of systematic uncertainty are discussed below. Unless
specified otherwise, each source of systematic uncertainty is varied
by ${\pm}1$ standard deviation
to infer the relative variation in
each of the quantities appearing in the likelihood function.

The uncertainty in the simulated jet energy
scale depends on the jet \pt and $\eta$, and is smaller than
3\%~\cite{Chatrchyan:2011ds}. This uncertainty is also propagated to
the simulated \MET calculation. It induces
uncertainties of the order of 10\% in the efficiency values and in the
W/Z+jets contributions to $N^\mathrm{B}_{\text{bck}}$ and $N^\mathrm{T}_{\text{bck}}$,
but its impact on the final limits remains
very limited because of the highly correlated effects on these quantities.

The jet energy resolution is varied in the simulation within its uncertainty,
which is of the order of 10\%~\cite{Chatrchyan:2011ds}. This
uncertainty is also propagated into the simulated \MET
calculation. Although it induces a relative change in the efficiency
values and in the W/Z+jets yield of less than 5\%, it is one of the
sources of uncertainty with the largest impact on the final
limits. In fact assuming no uncertainty in the jet energy resolution
causes  the expected upper limit at 95\% CL on $\mathcal{B}$ to
decrease by about 15\%.

The  uncertainty in the yield of the QCD multijet background is
discussed in Section~\ref{sec:bgestimate2}.
This uncertainty has a
significant impact only in the electron analysis, where its effect is
comparable to that of the jet energy resolution uncertainty.

The uncertainties in the cross section values of the W+jets and Z+jets
backgrounds largely dominate the uncertainty in the values of
$N^\mathrm{B}_{\text{bck}}$ and $N^\mathrm{T}_{\text{bck}}$ for
the muon analysis, whereas in the electron analysis they are
comparable to the uncertainty in the QCD multijet contribution. As described in
Section~\ref{sec:bgestimate}, the W+jets and Z+jets cross section
values used in this analysis are the theoretical inclusive predictions,
which have an uncertainty of about 5\%~\cite{WZcross}.
The CMS measurement~\cite{Chatrchyan:2011ne} of the ratio of the
W$+4\,$jets to the inclusive W cross section (the result for
W$+5\,$jets is not available) is in agreement with the \MADGRAPH
predictions within the measurement uncertainty, which is at the level
of 30\%.
In addition,
the limited number of events in the
simulated W+jets and Z+jets
samples introduces a statistical uncertainty of about 10\% in the
yield of these processes in the tight selection.
Taking these contributions into account, a conservative uncertainty of
50\% in the W+jets and Z+jets cross sections is assumed. This
uncertainty is found to have an impact comparable to that of the jet
energy resolution on the final limits.

The uncertainties in the final results, related to the  factorization
and renormalization scales
and to the  matching thresholds used for interfacing the matrix
elements generated with \MADGRAPH and the \PYTHIA parton showering,
are evaluated with dedicated simulated data samples where the nominal
values of the thresholds or scales are halved or doubled.
The uncertainty in the scales is found to have an impact on the final
limits comparable to that of the jet energy resolution 
uncertainty, while the impact of the uncertainty in the matching
thresholds is almost a factor of two smaller than that of the jet 
energy resolution.
The simulated samples are generated using the CTEQ 6.6
parton distribution functions~\cite{PhysRevD.78.013004}. The impact of
the uncertainties in the parton distribution functions is studied
following the PDF4LHC prescription~\cite{Botje:2011sn, Alekhin:2011sk,
Lai:2010vv, Martin:2009iq, Ball:2011mu} and is found to be very close
to that of the matching thresholds.

A number of other sources of systematic uncertainties are found to have
a negligible impact on the final results. They are summarized in the
following. The uncertainty in the efficiency of the lepton trigger,
identification, and isolation is assessed to be  5\%~\cite{Chatrchyan:2012xi,
Chatrchyan:2013dga} for both muons and electrons.
Unclustered reconstructed particles are also used to
compute \MET\ ~\cite{Chatrchyan:2011tn}, and thus an uncertainty in the
model provided by simulation of these particles is
reflected in an uncertainty in the final \MET calculation. The associated
systematic uncertainty  is estimated by varying the
contribution of unclustered particles to \MET by ${\pm}10\%$.
An uncertainty of 5\% in the estimated mean number of pileup
collisions is assumed.
An uncertainty of 2.6\% is assigned to the integrated
luminosity~\cite{CMS-PAS-LUM-13-001}.
This uncertainty affects
only $N^\mathrm{B}_{\text{bck}}$ and $N^\mathrm{T}_{\text{bck}}$,
but is negligible compared with other sources of uncertainty.
The uncertainty in the b-tagging efficiency results in an uncertainty
in the event selection efficiency in the range 1\% to 5\%
depending on the number, energy, $\eta$, and type of the jets in the
event~\cite{Chatrchyan:2012jua}. The consequent uncertainty induced in
$N^\mathrm{T}_{\text{exp}}$ is about 3\% for both the muon and
electron channels.
The uncertainties in the \ttbar and tW production cross section
values are about 5\%~\cite{PhysRevLett.110.252004} and
7\%~\cite{xsec_ttbar} arising from the uncertainties in the
factorization and renormalization scales, the parton distribution
functions and, in the case of the \ttbar cross section, the top-quark mass.
 These two uncertainties affect the ratio
$\sigma_{\cPqt\PW}/\sigma_{\ttbar}$ in Eq.~(\ref{eq:BR}) and are
conservatively assumed to be uncorrelated. 
The uncertainties in the yields from the WW, WZ, and ZZ processes, as
well as from $s$- and $t$-channel single-top-quark production are neglected
since these processes make only small contributions to
$N^\mathrm{B}_{\text{bck}}$ and $N^\mathrm{T}_{\text{bck}}$.

The central values and the overall uncertainties in the quantities used for
the calculation of the likelihood function are reported in
Table~\ref{tab::nuisance}
for both the muon and electron analyses.
\begin{table}[htb]
\begin{center}
\topcaption{\label{tab::nuisance} Central values and associated overall
uncertainties for the quantities appearing in the
likelihood function.}
\setlength\extrarowheight{2pt}
\begin{tabular}{ l | c  c }
Quantity  & Muon channel & Electron channel \\ \hline
$\epsilon^\mathrm{B}_{\mathrm{SM,SM}}$    & $(8.1 \pm 1.5) \times 10^{-3}$ & $(8.0 \pm 1.5) \times 10^{-3}$ \\
$\epsilon^\mathrm{T}_{\mathrm{SM,SM}}$    & $(4.62 \pm 0.93) \times 10^{-4}$ & $(4.24 \pm 0.85) \times 10^{-4}$ \\
$\epsilon^\mathrm{B}_{\mathrm{BNV,SM}}$   & $(7.37 \pm 0.89) \times 10^{-2}$ & $(7.33 \pm 0.88) \times 10^{-2}$ \\
$\epsilon^\mathrm{T}_{\mathrm{BNV,SM}}$   & $(1.86 \pm 0.32) \times 10^{-2}$ & $(1.62 \pm 0.27) \times 10^{-2}$ \\
$\epsilon^\mathrm{B}_{\mathrm{BNV,BNV}}$  & $(1.00 \pm 0.16) \times 10^{-2}$ & $(1.55 \pm 0.25) \times 10^{-2}$  \\
$\epsilon^\mathrm{T}_{\mathrm{BNV,BNV}}$  & $(1.74 \pm 0.32) \times 10^{-3}$ & $(2.64 \pm 0.55) \times 10^{-3}$ \\
$\epsilon^\mathrm{B}_{\mathrm{SM}}$     & $(2.68 \pm 0.32) \times 10^{-3}$ & $(2.57 \pm 0.31) \times 10^{-3}$ \\
$\epsilon^\mathrm{T}_{\mathrm{SM}}$     & $(1.13 \pm 0.14) \times 10^{-4}$ & $(8.21 \pm 0.99) \times 10^{-5}$ \\
$\epsilon^\mathrm{B}_{\mathrm{BNV}}$    & $(2.72 \pm 0.42) \times 10^{-2}$ & $(2.80 \pm 0.42) \times 10^{-2}$ \\
$\epsilon^\mathrm{T}_{\mathrm{BNV}}$    & $(5.38 \pm 0.84) \times 10^{-3}$ & $(5.84 \pm 0.82) \times 10^{-3}$  \\
$N^\mathrm{B}_{\text{bck}}$                      & $8100 \pm 3400$ & $10600 \pm 3700$ \\
$N^\mathrm{T}_{\text{bck}}$                      & $390 \pm 140$ &   $680 \pm 230$ \\

$N^\mathrm{B}_{\text{obs}}$                      & $47950 \pm 220$ & $50110 \pm 220$ \\
$\sigma_{\cPqt\PW}$                    & \multicolumn{2}{c}{$22.2 \pm 1.5$ pb}  \\
$\sigma_{\ttbar}$                & \multicolumn{2}{c}{$246 \pm 12$ pb}  \\
\end{tabular}
\end{center}
\end{table}

\section{Results \label{sec:results}}

Tables~\ref{tab::basictightnoBR_Wttmeasured}
and~\ref{tab::basictightnoBR_WttmeasuredEle} report the  yields
expected from the different SM processes considered,
and the yields observed in data for the muon and electron
channels, respectively.
In these tables $\mathcal{B}$ is assumed to be zero.
The yields of the \ttbar and tW processes in the basic selection
before and after the normalization procedure described in
Section~\ref{sec:brmeasurement} are both reported, the yields before
normalization being simply calculated as the products of the
theoretical cross sections, the measured values of the integrated luminosity,
and the event selection efficiencies obtained from simulation.
Because of the normalization procedure many of the uncertainties
reported in Tables~\ref{tab::basictightnoBR_Wttmeasured}
and~\ref{tab::basictightnoBR_WttmeasuredEle} are correlated or
anticorrelated, which explains why the uncertainties in the total
expected yields are smaller than those in some of their components.
In both the muon and electron channels the observed total yield in the
tight selection agrees with the SM expectations.

\begin{figure*}[htbp]
\begin{center}
\includegraphics[width=0.48\textwidth]{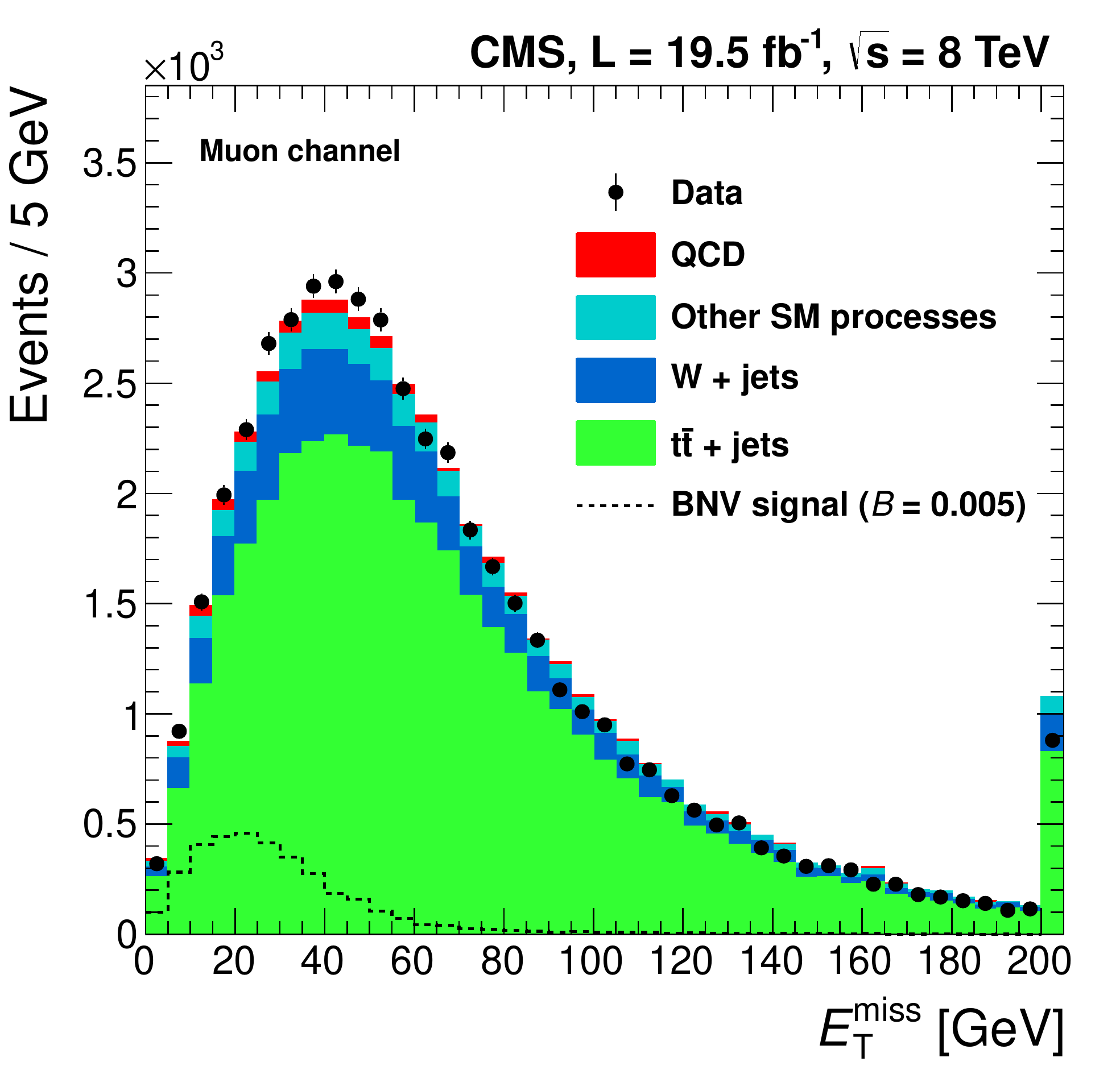}
\includegraphics[width=0.48\textwidth]{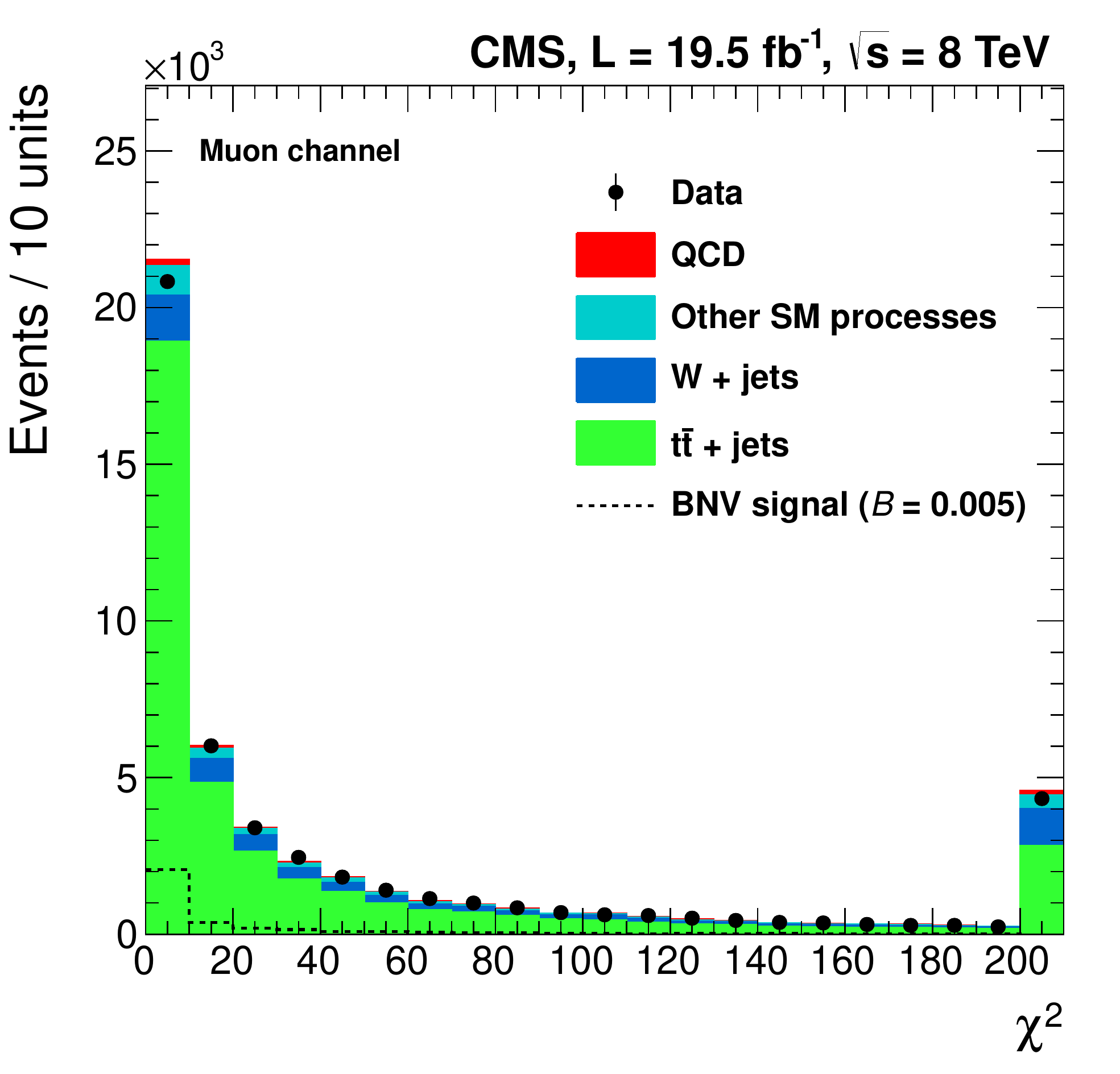}
\includegraphics[width=0.48\textwidth]{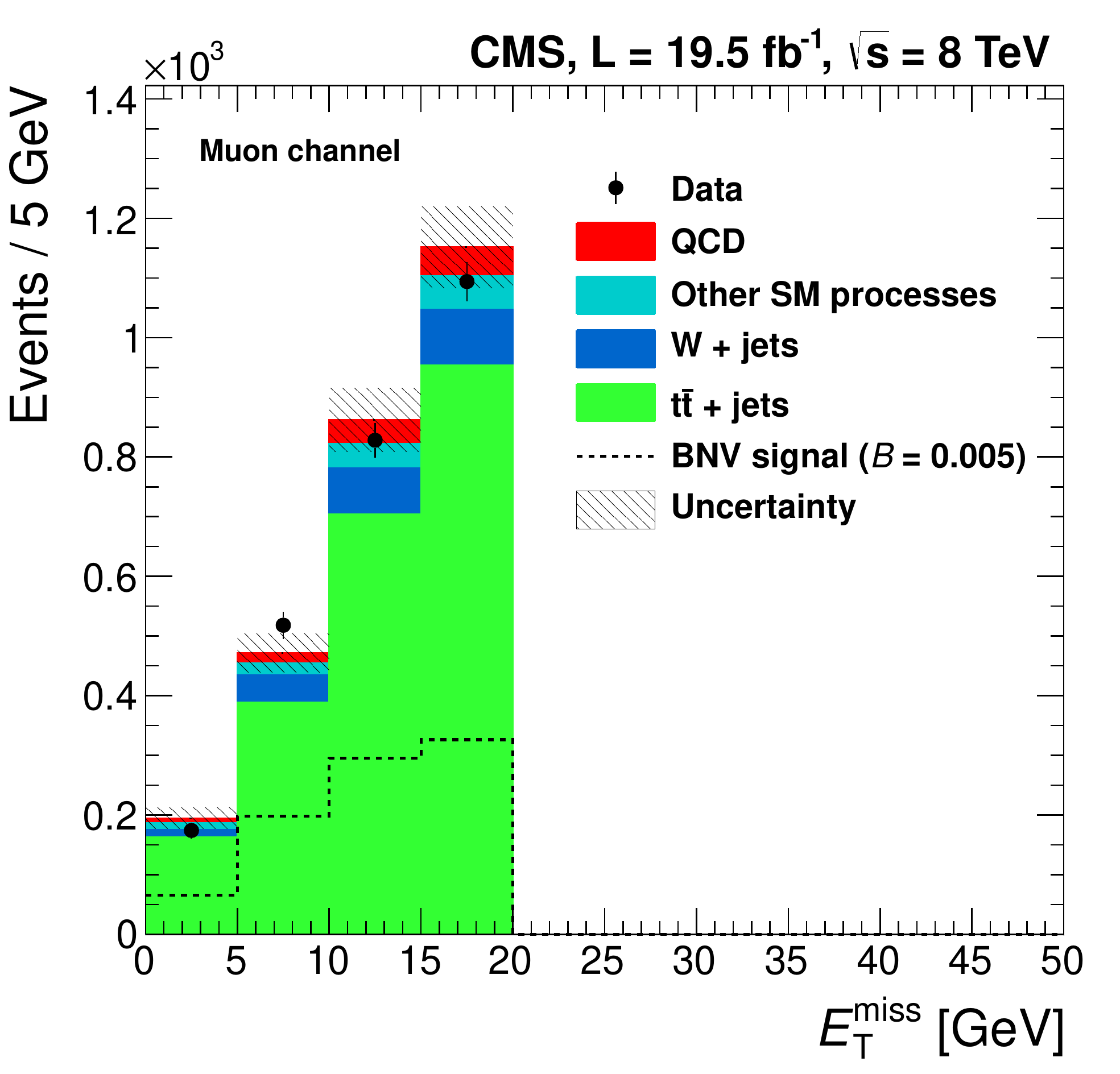}
\includegraphics[width=0.48\textwidth]{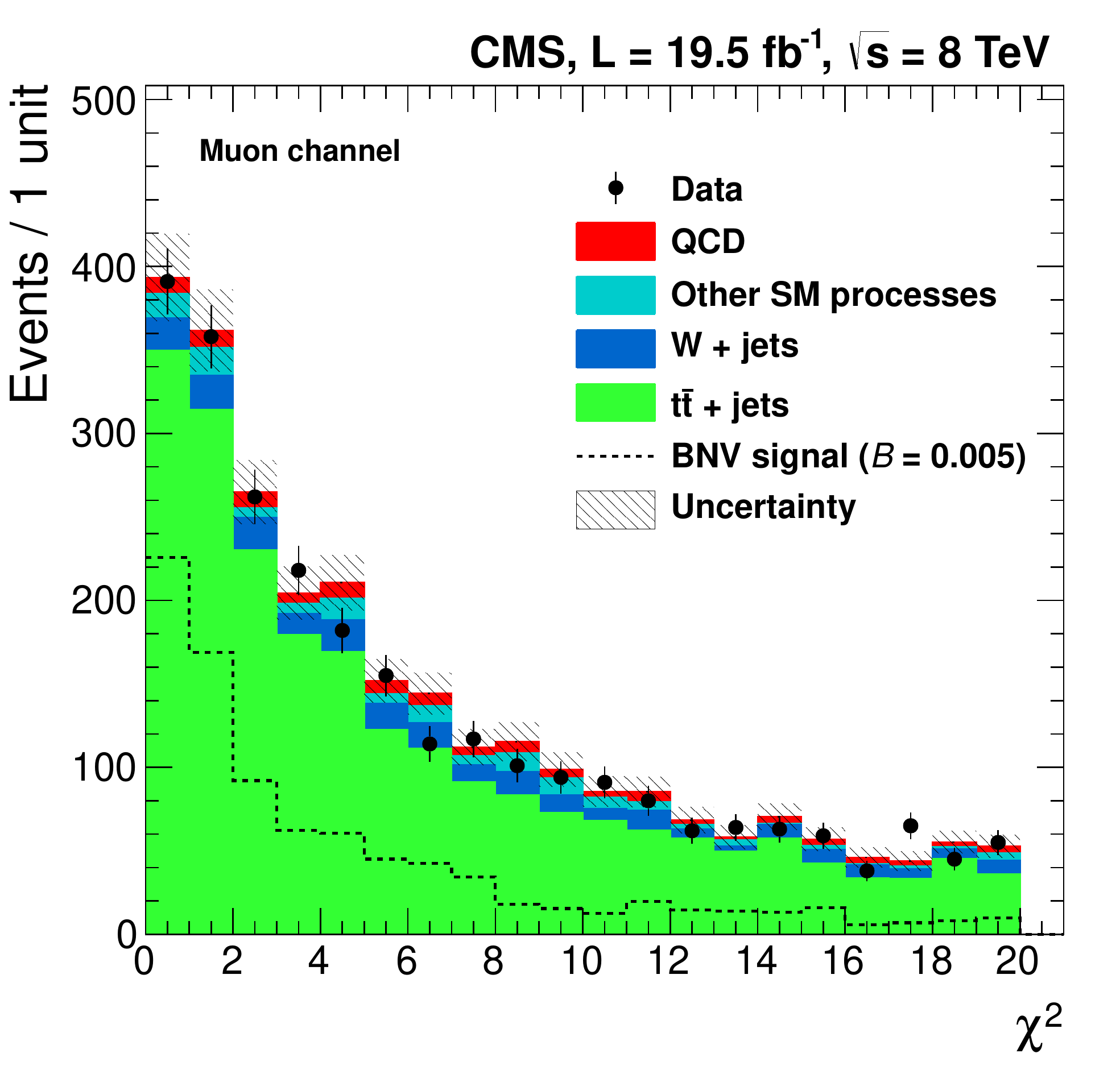}
\caption{Muon channel: observed and SM expected distributions of \MET
(left) and $\chi^2$ (right). The signal contribution expected for
a branching fraction $\mathcal{B}=0.005$ for the baryon number violating
top-quark decay is also shown.  Top: distributions for the basic
selection; because of the normalization to data, the integrals of the two
distributions are equal; overflowing entries are included in the
last bins of the distributions.
Bottom:  distributions for the tight
selection; the shaded band indicating the total uncertainty in the expected
yield is estimated assuming that the systematic relative uncertainty has no
dependence on  \MET or $\chi^2$.
}
\label{fig:muondist}
\end{center}
\end{figure*}

\begin{figure*}[htbp]
\begin{center}
\includegraphics[width=0.48\textwidth]{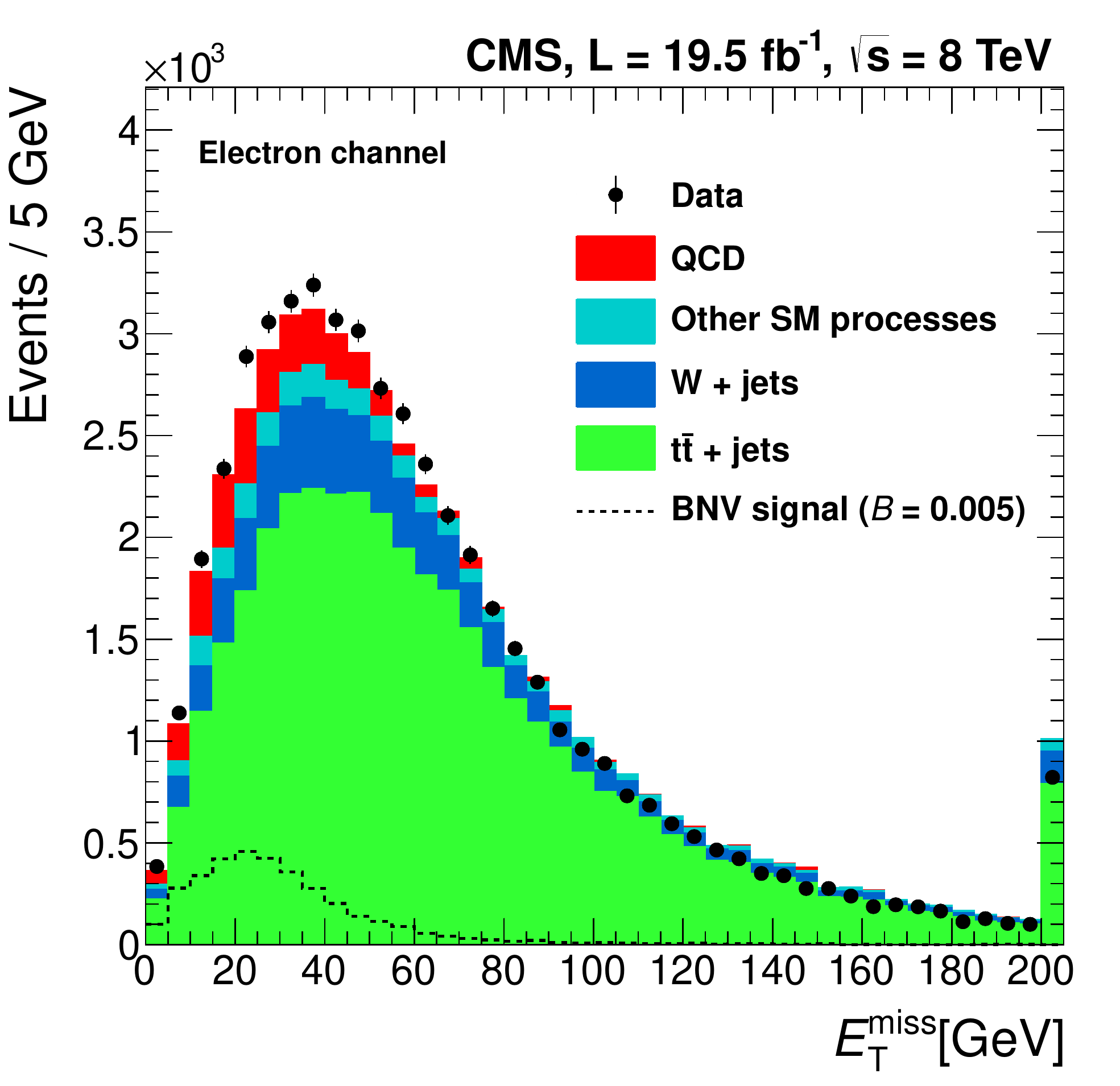}
\includegraphics[width=0.48\textwidth]{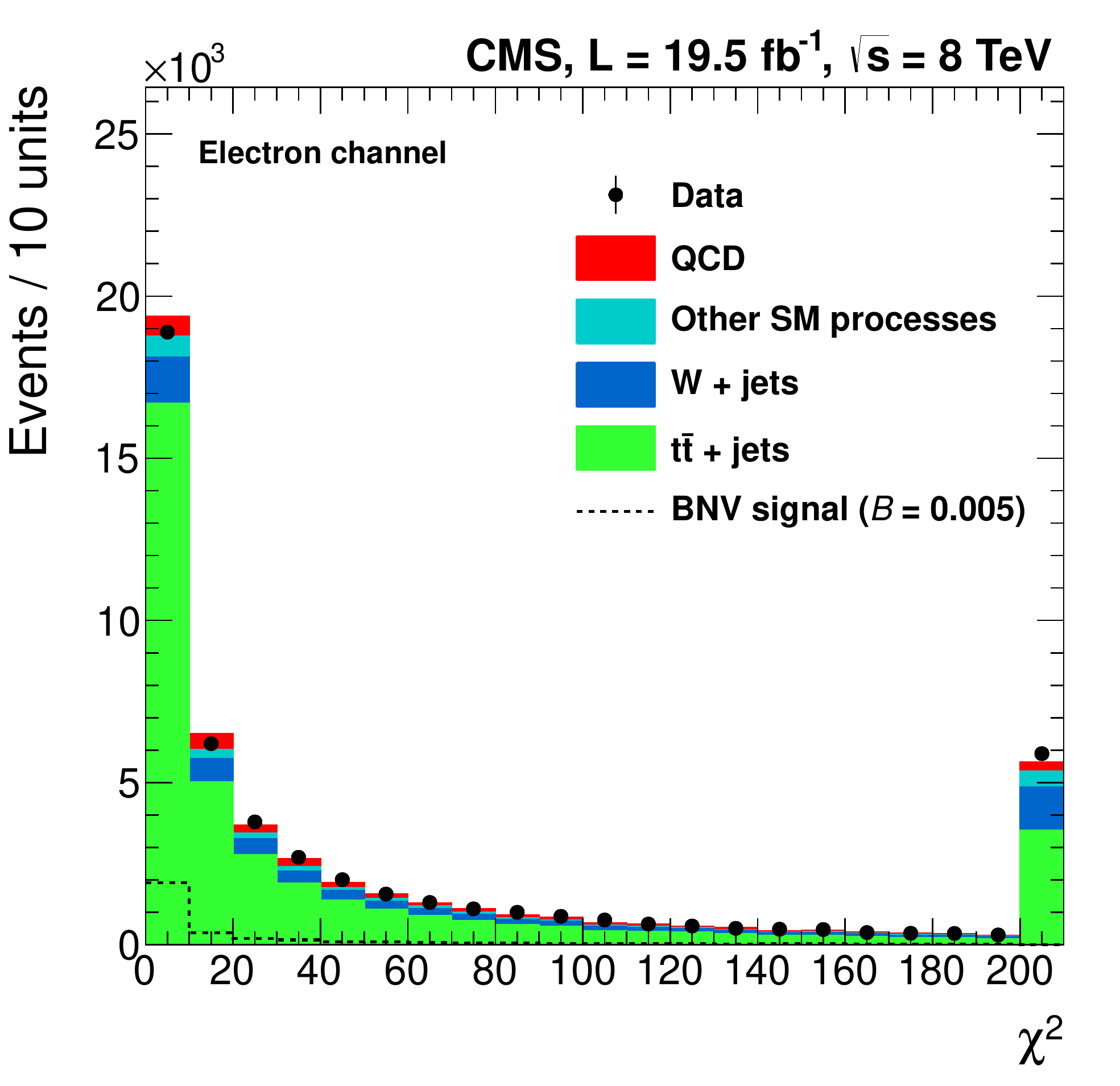}
\includegraphics[width=0.48\textwidth]{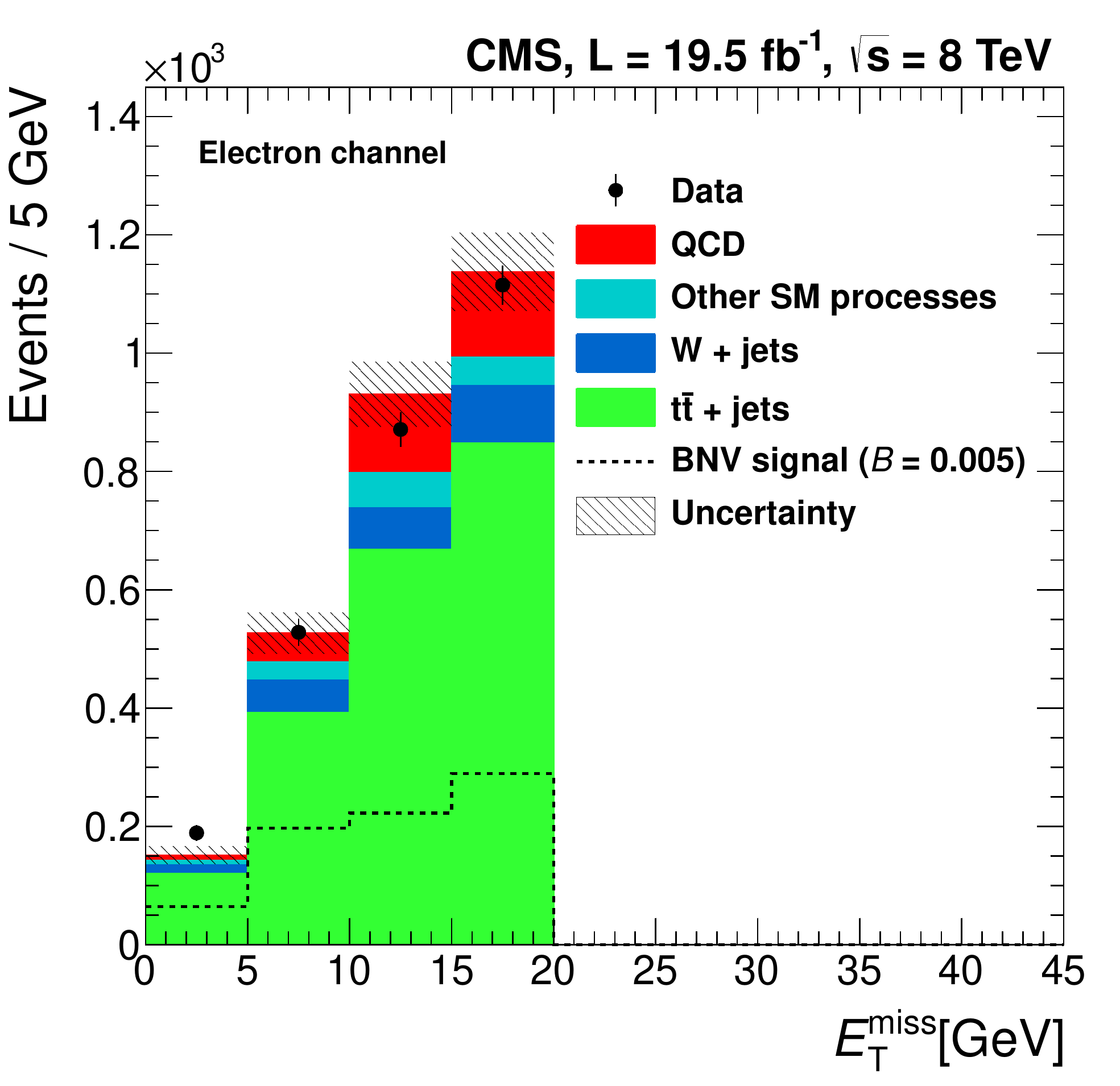}
\includegraphics[width=0.48\textwidth]{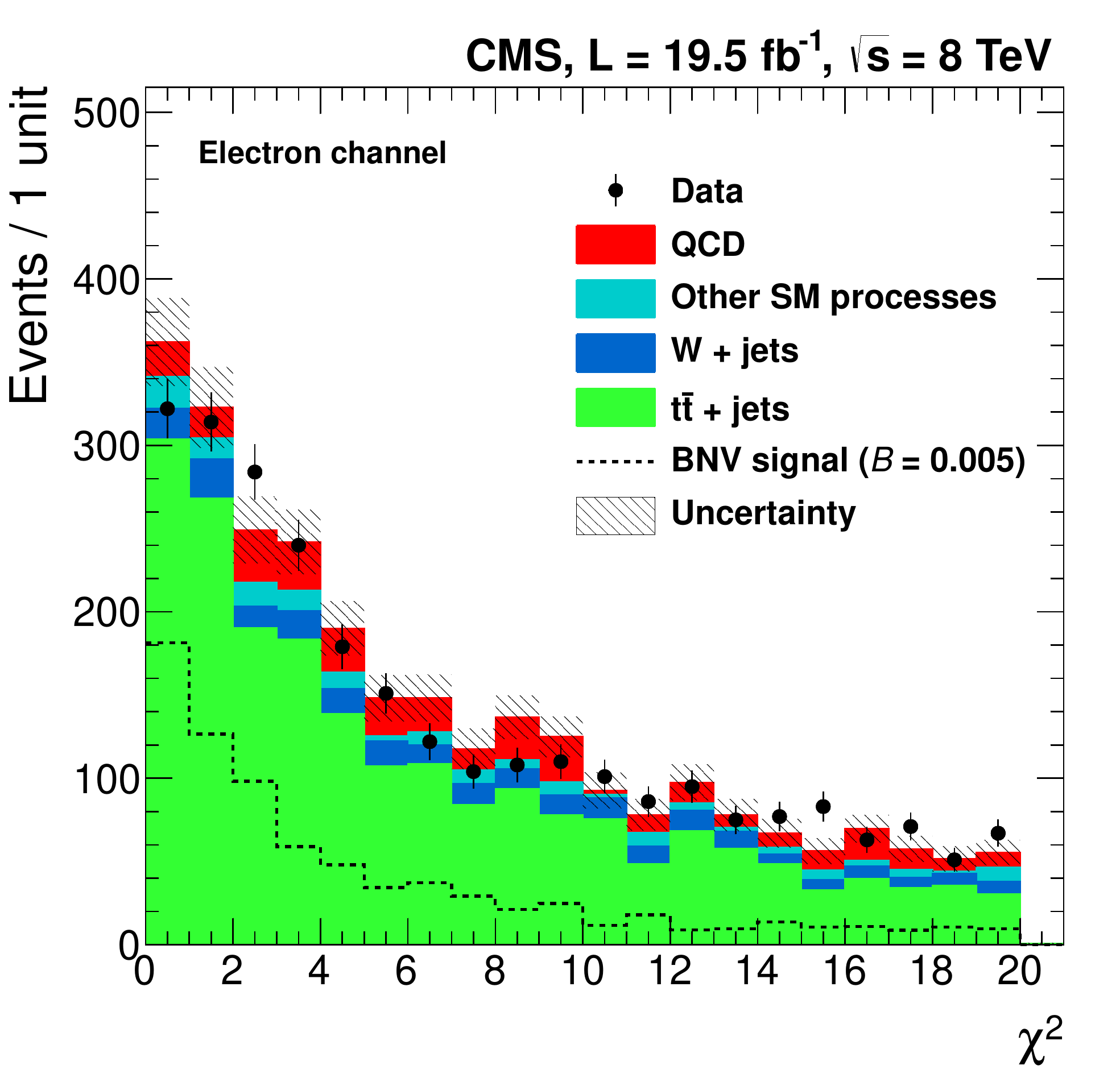}
\caption{Electron channel: observed and SM expected distributions of \MET
(left) and $\chi^2$ (right). The signal contribution expected for
a branching fraction $\mathcal{B}=0.005$ for the baryon number violating
top-quark decay is also shown. Top: distributions for the basic
selection; because of the normalization to data, the integrals of the two
distributions are equal; overflowing entries are included in the
last bins of the distributions.
Bottom:  distributions for the tight
selection;  the shaded band indicating the total uncertainty in the expected
yield is estimated assuming that the systematic relative uncertainty has no
dependence on \MET or $\chi^2$.
}
\label{fig:electrondist}
\end{center}
\end{figure*}

Figure~\ref{fig:muondist} (\ref{fig:electrondist})
shows, for the muon (electron) channel, the observed and expected
distributions of \MET and $\chi^2$
for the basic and tight selections assuming no BNV top-quark
decays. The signal distribution expected for $\mathcal{B}=0.005$ is also shown.
The methods adopted for the QCD multijet background estimates provide
no detailed shape information for this contribution. Thus, for the
sake of illustration in the plots, the shape of the QCD multijet
background contribution in the \MET distribution is obtained from
simulation using events with at least three jets, instead of five,
in order to reduce large statistical fluctuations.
In the case of  the $\chi^2$  variable, whose calculation requires events with
at least five jets, the shape of the QCD multijet background contribution is
obtained from data in the anti-isolated regions defined in
Section~\ref{sec:bgestimate2} after subtraction of the top-quark and
electroweak components estimated from simulation.
The discrepancy visible between the observed and expected \MET distributions
in the electron channel basic selection can be accommodated
with the 50\% uncertainty assumed in the total QCD multijet yield.
These distributions are presented for purposes of illustration and are
not used in the analysis.

The observed data samples are then used to calculate upper limits on
the value of $\mathcal{B}$.
The upper limit on $\mathcal{B}$ at 95\% CL is obtained with
the Feldman--Cousins approach~\cite{Feldman:1997qc}.
Pseudoexperiments are generated using the frequentist
prescription described in Ref.~\cite{Chatrchyan:2012tx}.
The results are summarized in Table~\ref{tab::resultslimit}
for the muon and electron channels, and  for their combination.
The combined results are obtained by
maximizing the product of the two likelihood functions, assuming a
common value of $\mathcal{B}$ for the two channels. Full correlation is
assumed for each pair of corresponding nuisance parameters in the two
analyses, except for those related to the lepton trigger,
identification, and isolation, which are assumed to be independent.
The combination of the muon and electron datasets does not
significantly improve the upper limit because of the dominant
systematic uncertainties related to the modeling of jets,
\MET, and event kinematic properties, which are fully correlated across the
two channels.
\begin{table}[htbp]
\begin{center}
\topcaption{\label{tab::resultslimit} Observed 95\% CL upper limit on
$\mathcal{B}$, expected median 95\% CL limit for the $\mathcal{B}=0$ hypothesis and
ranges that are expected to contain 68\% of all observed deviations
from the expected median  for the muon and electron
channels and for their combination.}
\begin{tabular}{ c  c  c  c }
 Channel & 95\% CL & Expected  & 68\% CL exp. range \\
 \hline
Muon  & 0.0016 & 0.0029 & [0.0017, 0.0046] \\
Electron& 0.0017 & 0.0030 & [0.0017, 0.0047] \\
Combined  & 0.0015 & 0.0028 & [0.0016, 0.0046]  \\
\end{tabular}
\end{center}
\end{table}

\section{Summary}
Data recorded by the CMS detector have been used
to search for baryon number violation in top-quark decays.
The data correspond to an integrated luminosity of $19.52 \pm
0.49\fbinv$ at $\sqrt{s} = 8$\TeV.
No significant excess is observed over the SM expectation
for events with one isolated lepton (either a muon or an electron),
at least five jets of which at least one is b tagged, and low missing
transverse energy.
These results are used to set an upper limit of 0.0016~(0.0017)
at 95\% confidence level on the branching
fraction of a hypothetical baryon number violating top-quark decay into a
muon (electron) and 2 jets. The combination of the two channels under
the assumption of lepton universality yields an upper limit of 0.0015.
These limits on baryon number violation are the
first that have been obtained for a process
involving the top quark.

\section*{Acknowledgements}

We congratulate our colleagues in the CERN accelerator departments for the excellent performance of the LHC and thank the technical and administrative staffs at CERN and at other CMS institutes for their contributions to the success of the CMS effort. In addition, we gratefully acknowledge the computing centres and personnel of the Worldwide LHC Computing Grid for delivering so effectively the computing infrastructure essential to our analyses. Finally, we acknowledge the enduring support for the construction and operation of the LHC and the CMS detector provided by the following funding agencies: BMWF and FWF (Austria); FNRS and FWO (Belgium); CNPq, CAPES, FAPERJ, and FAPESP (Brazil); MES (Bulgaria); CERN; CAS, MoST, and NSFC (China); COLCIENCIAS (Colombia); MSES (Croatia); RPF (Cyprus); MoER, SF0690030s09 and ERDF (Estonia); Academy of Finland, MEC, and HIP (Finland); CEA and CNRS/IN2P3 (France); BMBF, DFG, and HGF (Germany); GSRT (Greece); OTKA and NKTH (Hungary); DAE and DST (India); IPM (Iran); SFI (Ireland); INFN (Italy); NRF and WCU (Republic of Korea); LAS (Lithuania); CINVESTAV, CONACYT, SEP, and UASLP-FAI (Mexico); MBIE (New Zealand); PAEC (Pakistan); MSHE and NSC (Poland); FCT (Portugal); JINR (Dubna); MON, RosAtom, RAS and RFBR (Russia); MESTD (Serbia); SEIDI and CPAN (Spain); Swiss Funding Agencies (Switzerland); NSC (Taipei); ThEPCenter, IPST, STAR and NSTDA (Thailand); TUBITAK and TAEK (Turkey); NASU (Ukraine); STFC (United Kingdom); DOE and NSF (USA).

Individuals have received support from the Marie-Curie programme and the European Research Council and EPLANET (European Union); the Leventis Foundation; the A. P. Sloan Foundation; the Alexander von Humboldt Foundation; the Belgian Federal Science Policy Office; the Fonds pour la Formation \`a la Recherche dans l'Industrie et dans l'Agriculture (FRIA-Belgium); the Agentschap voor Innovatie door Wetenschap en Technologie (IWT-Belgium); the Ministry of Education, Youth and Sports (MEYS) of Czech Republic; the Council of Science and Industrial Research, India; the Compagnia di San Paolo (Torino); the HOMING PLUS programme of Foundation for Polish Science, cofinanced by EU, Regional Development Fund; and the Thalis and Aristeia programmes cofinanced by EU-ESF and the Greek NSRF.

\bibliography{auto_generated}   

\cleardoublepage \appendix\section{The CMS Collaboration \label{app:collab}}\begin{sloppypar}\hyphenpenalty=5000\widowpenalty=500\clubpenalty=5000\textbf{Yerevan Physics Institute,  Yerevan,  Armenia}\\*[0pt]
S.~Chatrchyan, V.~Khachatryan, A.M.~Sirunyan, A.~Tumasyan
\vskip\cmsinstskip
\textbf{Institut f\"{u}r Hochenergiephysik der OeAW,  Wien,  Austria}\\*[0pt]
W.~Adam, T.~Bergauer, M.~Dragicevic, J.~Er\"{o}, C.~Fabjan\cmsAuthorMark{1}, M.~Friedl, R.~Fr\"{u}hwirth\cmsAuthorMark{1}, V.M.~Ghete, N.~H\"{o}rmann, J.~Hrubec, M.~Jeitler\cmsAuthorMark{1}, W.~Kiesenhofer, V.~Kn\"{u}nz, M.~Krammer\cmsAuthorMark{1}, I.~Kr\"{a}tschmer, D.~Liko, I.~Mikulec, D.~Rabady\cmsAuthorMark{2}, B.~Rahbaran, C.~Rohringer, H.~Rohringer, R.~Sch\"{o}fbeck, J.~Strauss, A.~Taurok, W.~Treberer-Treberspurg, W.~Waltenberger, C.-E.~Wulz\cmsAuthorMark{1}
% ...
\vskip\cmsinstskip
% ...
1:~~Also at Vienna University of Technology, Vienna, Austria\\
2:~~Also at CERN, European Organization for Nuclear Research, Geneva, Switzerland\\
% ...

\end{sloppypar}
\end{document}